\documentclass{article}

\usepackage{arxiv}

\usepackage[utf8]{inputenc} 
\usepackage[T1]{fontenc}    
\usepackage{hyperref}       
\usepackage{url}            
\usepackage{booktabs}       
\usepackage{amsfonts}       
\usepackage{nicefrac}       
\usepackage{microtype}      
\usepackage{lipsum}		
\usepackage{graphicx}
\usepackage{natbib}
\usepackage{doi}

\usepackage{amsmath}
\usepackage{amssymb}
\usepackage{cool}
\usepackage{colortbl}%

\newcommand{\RRSet}{{\mathrm I\! \mbox{R} }}

\newcommand{\va}{{\mathbf a}}
\newcommand{\vb}{{\mathbf b}}

\newcommand{\vc}{{\mathbf c}}
\newcommand{\vd}{{\mathbf d}}

\newcommand{\vr}{{\mathbf r}}

\newcommand{\vs}{{\mathbf s}}

\newcommand{\vv}{{\mathbf v}}
\newcommand{\vw}{{\mathbf w}}

\newcommand{\vx}{{\mathbf x}}

\newcommand{\vy}{{\mathbf y}}

\newcommand{\invn}[2][0cm]{\mathopen{}\left|\left|{#2}\parbox[h][#1]{0cm}{}\right|\right|}

\newcommand{\mC}{{\mathbf C}}

\newcommand{\mI}{{\mathbf I}}
\newcommand{\mJ}{{\mathbf J}}

\newcommand{\mR}{{\mathbf R}}

\newcommand{\NNull}{{\mathbf 0}}

\Style{IdentityMatrixSymb=\mI,%
       DSymb={\mathrm d},%
       IntegrateDifferentialDSymb={\mathrm d}}

\newtheorem{definition}{Definition}

\newtheorem{corollary}{Corollary}

\usepackage{tikz}
\usetikzlibrary{positioning}

\title{Scalar Representation of 2D Steady Vector Fields}

\author{ \href{https://orcid.org/0000-0002-8009-7070}{\includegraphics[scale=0.06]{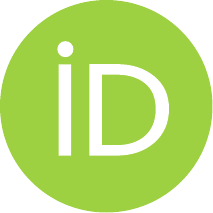}\hspace{1mm}Holger Theisel}
		\\
	Department of Computer Science\\
	University of Magdeburg\\
	Magdeburg, Germany\\
	\texttt{theisel@ovgu.de} \\
	\And
	{\includegraphics[scale=0.06]{orcid.pdf}\hspace{1mm}Michael Motejat} \\
	Department of Computer Science\\
	University of Magdeburg\\
	Magdeburg, Germany\\
	\texttt{michael@isg.cs.uni-magdeburg.de} \\
	\And
	{\includegraphics[scale=0.06]{orcid.pdf}\hspace{1mm}Janos Zimmermann} \\
	Department of Computer Science\\
	University of Magdeburg\\
	Magdeburg, Germany\\
	\texttt{janos@isg.cs.uni-magdeburg.de} \\
	\And
	{\includegraphics[scale=0.06]{orcid.pdf}\hspace{1mm}Christian R\"ossl} \\
	Department of Computer Science\\
	University of Magdeburg\\
	Magdeburg, Germany\\
	\texttt{roessl@isg.cs.uni-magdeburg.de} \\
}

\renewcommand{\headeright}{}
\renewcommand{\undertitle}{}

\begin{document}
\maketitle

\begin{abstract}
We introduce a representation of a 2D steady vector field $\vv$ by two scalar fields $a$, $b$, such that the isolines of $a$ correspond 
to stream lines of $\vv$, and $b$ increases with constant speed  under integration of $\vv$. This way, we get a direct encoding of stream lines, i.e., a numerical integration of $\vv$ can be replaced by a local isoline extraction of $a$. To guarantee a solution in every case, gradient-preserving cuts are introduced such that the scalar fields are allowed to be discontinuous in the values but continuous in the gradient. Along with a piecewise linear discretization and a proper placement of the cuts, the fields $a$ and $b$ can be computed. We show several evaluations on non-trivial vector fields.
\end{abstract}


\section{Introduction}

Scalar fields and vector fields are perhaps the most common data classes in (Scientific) Visualization. For both classes, a huge amount of visualization techniques has been developed over the last decades. Vector fields are usually the more complicated data class because firstly they contain more data  per point in the domain and secondly transportation issues of particles are involved.  Because of this, scalar field visualization is much further developed than vector field visualization.

A general approach to vector field visualization is to derive one (ore more) scalar fields from a vector field, and then apply scalar field visualization on them. While many scalar fields have been proposed to visualize vector fields (like vector magnitude, divergence, FTLE,...), all of them come with a loss of information. In particular, they do not encode the transport of particles, i.e., the location of particles after a certain integration time cannot be computed (neither directly nor indirectly by numerical integration) from these scalar fields.

Flow visualization has a variety of goals and applications, ranging from understanding of fundamental flow phenomena to the analysis of concrete simulations or measurements. We identify generic problems that frequently occur in various applications of flow visualization:
\begin{itemize}
\item
{\em The integration problem:} many algorithms in vector field visualization are based on the numerical integration of stream lines / particle trajectories. While stream line integration is numerically well-understood, it is still source of error because of error accumulation during the integration. Error accumulation is an issue for every numerical stream lines integration technique, no matter how involved it is.
\item
{\em The connectivity problem:} many algorithms rely on an efficient approach to answer the following question: given two points $\vx_1,\vx_2$, does a streamline starting from $\vx_1$ hit the point $\vx_2$? If not, on which side (in 2D) and at which distance the line passes by? 
 \end{itemize}
These two problems are generic: they appear in many visualization techniques. Having efficient solutions for them would affect different existing visualization techniques.

In this paper, we follow the established path of finding scalar fields representing vector fields. However, our new approach is to find scalar fields that solve both the integration problem and the connectivity problem. In other words: we search for derived scalar fields such that both integration and connectivity can be solved by a local lookup, a procedure much faster and less error prone than numerical stream line integration. 
 In particular, we want to find a scalar field such that the stream lines of the vector field correspond to isolines of the scalar field. 
 
 Unfortunately, in general a scalar field with isolines following stream lines of vector fields does not exist. To overcome this, we weaken  our goal by allowing gradient-preserving cuts and  a special treatment of critical points. With this, we find scalar representations of vector fields that directly encode particle transportation. We apply our approach to introduce a strictly local  (i.e., without any numerical integration) image based flow visualization technique.

\subsection*{Notation}

We consider a 2D steady vector field  $\vv(\vx)$ over a simple connected limited domain $D \subset \RRSet^2$ with disk topology where $\delta D$ is the boundary curve of $D$. We assume that $\delta D$ is given as a closed differentiable parametric curve $\vd(s)$.
Let $\phi(\vx,\tau)$ denote the flow map of $\vv$, i.e., $\phi(\vx,\tau)$ is the location to be landed by integrating $\vv$ starting from $\vx$ over an integration time $\tau$. Let $\mJ$ be the Jacobian matrix of $\vv$, and let
\begin{eqnarray}
\overline{\vv}(\vx) = \frac{\vv}{\invn{\vv}}
\;,\;
\nonumber
\overline{\vw}(\vx)=
\begin{pmatrix}
  0 & -1 \\
  1 & 0
\end{pmatrix} \cdot \overline{\vv}
\end{eqnarray}
be the normalized and normalized perpendicular field of $\vv$.
Further, $\nabla a$ denotes the gradient of a scalar field $a$, while  $\nabla_\perp a = \begin{pmatrix}
  0 & -1 \\
  1 & 0
\end{pmatrix} 
\nabla a$ denotes its co-gradient.

\section{Problem description and analysis}

Given a 2D steady vector field $\vv(\vx)$, 
we search for two smooth and differentiable scalar fields $a(\vx)$, $b(\vx)$ with the following properties:
\begin{eqnarray}
\label{eq_cond_a_b_1}
\nabla a &\neq& \NNull \\
\label{eq_cond_a_b_2}
\vv^T \nabla a &=& 0\\
\label{eq_cond_a_b_3}
\vv^T \nabla b &=& 1
\end{eqnarray}
for all $\vx \in D$. (\ref{eq_cond_a_b_1}) and (\ref{eq_cond_a_b_2}) make sure that a stream line of $\vv$ is an isoline of $a$. (\ref{eq_cond_a_b_3}) makes sure that $b$ increases with unit speed under integrating a stream line. This way, stream line integration in $\vv$ reduces to isocurve intersection of $a$ and $b$ by
\begin{equation}
\label{eq_cond_a_b_4}
a(\phi(\vx,\tau)) = a(\vx)  \; \; \;,\; \; \; b(\phi(\vx,\tau)) = b(\vx) + \tau.
\end{equation}
Unfortunately, a field $a$ fulfilling (\ref{eq_cond_a_b_1}), (\ref{eq_cond_a_b_2}) does in general not exist. The simplest counterexample is the linear field $\vv(\vx)=(x,y)^T$ for which no smooth field $a$ can exist. In general, if $\vv$ is divergence-free, a field $a$ exists: the stream function. Also for the field $b$, existence is not ensured: for divergence-free  $\vv$,  a contradiction in $b$ is created on a closed stream line after one turn.    

In order to cope with the problem, we allow a certain weakening of the conditions for $a, b$ that allow a unique solution but still allow to solve the integration and connectivity problem by an isovalue lookup. In particular, we introduce two concepts: gradient preserving cuts and a special treatment of areas around critical points.

\section{Related work}
\label{sec_relatedwork}

\noindent
We divide the review of existing work in four parts: vector field visualization, scalar field visualization, scalar field representation of vector fields, and finding optimal cuts in shape processing.

\paragraph*{Vector field visualization}
\noindent
Vector field visualization is one of the core topics in Scientific Visualization. While in many cases the vector data describe flow phenomena, other  effects represented by vector field data exist as well, such as dynamical systems or magnetic fields. A variety of techniques for vector field visualization has been developed, ranging from texture based techniques \cite{Laramee2004} over feature extraction \cite{10.1111:j.1467-8659.2003.00723.x,Gun17f}, topological methods \cite{org:443:handle:10.1111:v30i6pp1789-1811,SJW*:2008} until illustrative techniques \cite{conf:EG2012:stars:075-094}. Most relevant for our work are topological techniques because they develop a partition of the vector fields in to critical points and separatrices of of relevance 
for the new approaches.

Topological methods for 2D vector fields have been introduced to
the visualization community in \cite{helman89}. Later they were
extended to higher order critical points \cite{scheuermann98},
boundary switch points \cite{deleeuw99}, and closed separatrices
\cite{wischgoll01}.
In addition, topological methods have been applied to
simplify \cite{deleeuw99, deleeuw99a, tricoche00, tricoche01a},
smooth \cite{westermann01},
compress \cite{lodha00,lodha03,theisel03ac}
and construct \cite{theisel02,weinkauf04b}
vector fields. 3D topological feature are considered in
\cite{globus91,helman91,mahrous03,mahrous04,theisel03c,weinkauf04a}.
State-of-the-Art-Reports on topological methods for flow visualization can be found in
\cite{laramee07,Pobitzer:2011:CGF,wang2016numerics,heine2016survey}.

Topological methods can be applied only to steady vector fields because they require integration over an infinitely long integration time. For unsteady fields, Lagrangian Coherent Structures (LCS) have been established to find regions of homogeneous flow behavior.
One of the most prominent approaches for this is the computation of ridge structures in FTLE fields, as introduced by Haller \cite{haller01,366505}. To consider spatial separation only, Pobitzer et al. \cite{pobitzer2012filtering} weighted FTLE values by their angle to the separation direction. FTLE ridges were proposed for a variety of applications \cite{haller02,lekien05,Shadden2009161,weldon08}. Shadden et al. \cite{shadden05} showed that ridges of FTLE are approximate material structures, i.e., they converge to material structures for increasing integration times. This fact was used in \cite{sadlo2009_timeTopo, uffinger2013time} to extract topological structures and in \cite{lipinski:017504}
to accelerate the FTLE computation in 2D flows.
Also in the visualization community, different approaches have been proposed to increase performance, accuracy and usefulness of FTLE as a visualization tool \cite{Guenther2016mcftl,SCI:Gar2007a,garth07,sadlo07b,sadlo07,rpbib:sadlo2009advection}.

In recent years approaches have evolved that aim at finding suitable moving frames of the underlying coordinate system to study the flow \cite{SDAV:Bha2014a,WGS:2007} or develop Galilean invariant detectors directly \cite{bujack2016topology}. This way, finite-time studies of time-dependent fields is lead back to a topological analysis of a derived steady field. 

We also note that discrete versions of vector field topology have been developed based on either Forman`s discrete Morse theory 
\cite{Forman01auser,ReininghausHotz2011}
or Morse decomposition \cite{Mischaikow2007}. Other approaches use edge maps \cite{DBLP:journals/tvcg/BhatiaJBCLNP12} or robustness considerations for critical points treatment
\cite{Skraba.2016.TVCG,DBLP:journals/cgf/WangRSBP13}.

\paragraph*{Scalar field visualization}
\noindent
Scalar fields are another well-researched standard data class in Scientific Visualization. A  variety of techniques has been developed for scalar fields, ranging from isosurface extraction over direct volume rendering to feature extraction. In particular, topological features have been proven successful. We mention Morse-Smale complexes and the Reeb graph.

The Morse-Smale complex is a topological structure  representing the gradient-flow behavior of a scalar field. Of particular interest for visualization are combinatorial approaches based on piecewise linear scalar fields as introduced in \cite{Edelsbrunner:2001:HMC:378583.378626} and later extended in several ways 
\cite{DBLP:journals/tvcg/BremerEHP04,EdelsbrunnerHNP03,
DBLP:conf/visualization/GyulassyNPBH05,DBLP:journals/tvcg/LewinerLT04,
DBLP:books/daglib/p/GyulassyBBP15,DBLP:journals/tvcg/GyulassyNPH07}. The combinatorial approach makes the extraction of Morse-Smale complexes relevant because it allows effective algorithms with guarantees about their correctness.

The Reeb graph considers components of contours and their topological changes. They have been applied to control the removal of topological features 
\cite{DBLP:conf/visualization/CignoniCMRS00,DBLP:conf/visualization/CarrSP04,
DBLP:conf/graphicsinterface/GuskovW01,DBLP:conf/gmp/TakahashiNTF04,
DBLP:journals/comgeo/CarrSA03,
DBLP:journals/tvcg/TiernyDNPS12}.

\paragraph*{Vector field approximation by scalar fields}
\noindent
Scalar fields are simpler structures than vector fields. So, it is not surprising that research came up with approaches to represent vector fields (in particular flow fields and vorticity fields) as scalar fields. The Clebsch map
\cite{Clebsch1859}
represents divergence-free  velocity and vorticity fields  as a certain combination of gradients of scalar fields. Such representations have been used to visualize 
\cite{Kotiuga91}, analyze \cite{jeong_hussain_1995} and simplify flows 
\cite{MNR:MNR15640,doi:10.1063/1.4943368,doi:10.1063/1.4943368}.
Unfortunately, such representations cannot be exact in the general case 
\cite{doi:10.1063/1.870331}, resulting in either approximate representations 
\cite{Chern:2016:SS}
or applications restricting to certain subclasses of divergence-free vector fields 
\cite{Angelidis:2007:KSU:1272690.1272709,vonFunck:2006:VFB:1141911.1142002}.

\paragraph*{Optimal cuts on shapes}
\noindent
Many algorithms in Geometry processing require open surfaces with well-defined boundaries, such as surface parametrization
\cite{zigelman02,Hormann:2008:MPT:1508044.1508091}
or quad-remeshing 
\cite{Ebke:2016:ICQ:2980179.2982413,10.1111:cgf.12401}.
For closed input surfaces, cutting algorithms are necessary. Since the placement of the cuts influences the result of the algorithms, "optimal" cuts are searched. Several criteria for optimality have been proposed and applied 
\cite{Erickson:2002:OCS:513400.513430}.

\section{Solution on simple vector fields}
\label{sec_solution_simple_field}

In this section, we provide a solution for $a$, $b$ for simple vector fields. 
\begin{definition}
A vector field $\vv$ to be {\em simple} if it fulfills the following property: for every $\vx \in D$ we reach the boundary after a finite integration time both in forward and backward integration. 
\end{definition}
From this it follows that $\vv$ has neither critical points nor closed orbits.

\begin{figure}
  \centering
    \includegraphics[width=0.5 \textwidth]{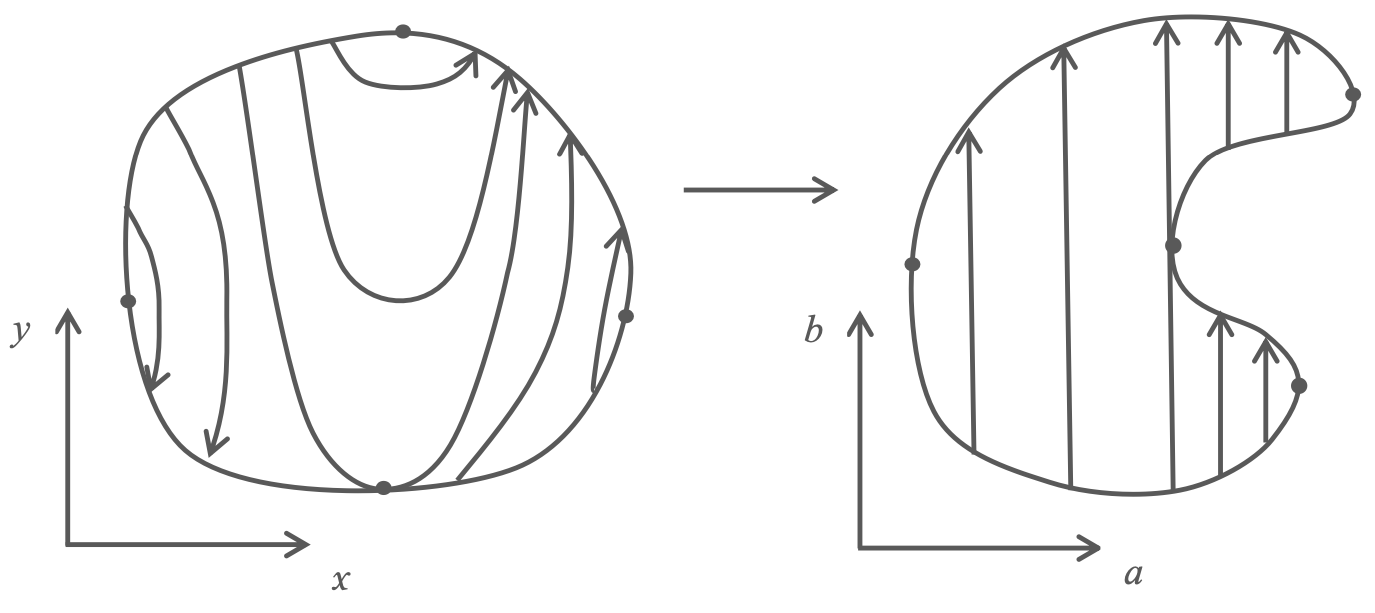}    \hfill
    \caption{Finding the fields $a,b$ can be interpreted as searching for a domain transformation of $\vv$ to $(0,1)^T$.}
  \label{fig_1}
 \end{figure}

\begin{figure}
  \centering
    \includegraphics[width=0.3 \textwidth]{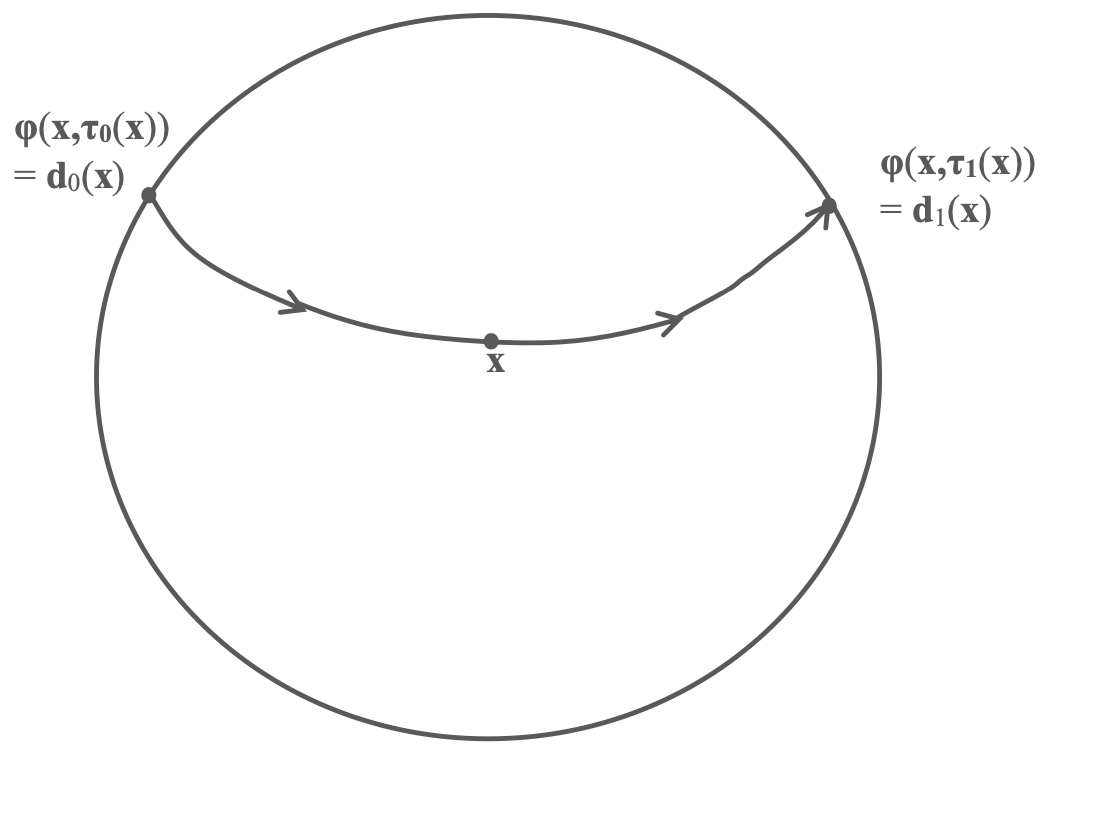}    \hfill
    \caption{Definition of $\tau_0(\vx),\tau_1(\vx)$, $\vd_0(\vx),\vd_1(\vx)$.}
  \label{fig_2}
 \end{figure}

For simple vector fields, the search for $a,b$ can be interpreted as a domain transformation of $\vv$ in the domain $D$ to the constant vector field $(0,1)^T$ in the new domain $D'$. Figure \ref{fig_1} gives an illustration.
Note that (\ref{eq_cond_a_b_1})--(\ref{eq_cond_a_b_3}) does not give unique solutions for $a,b$. Among all solutions for $a,b$, we search for the ones the with
\begin{eqnarray}
\int_D (\| \nabla a \| - \| \vv\| )^2 d \vx  &\to&  \min \\
\int_D \| \nabla b \|^2 d \vx &\to&  \min. 
\end{eqnarray}
Given a point $\vx \in D$, let $\tau_0(\vx)$, $\tau_1(\vx)$ be the integration time to reach the boundary of $D$ starting from $\vx$ in backward and forward direction respectively, i.e.:
\begin{eqnarray}
\tau_0(\vx) \leq 0 \leq \tau_1(\vx) \\
\phi(\vx,\tau_0(\vx)) , \phi(\vx,\tau_1(\vx)) \in \delta D\\
\phi(\vx,\tau) \notin  \delta D \mbox{   for   }   \tau  \in  \;  ] \tau_0(\vx),\tau_1(\vx)  [ .
\end{eqnarray}
Further, we define the end points of the integration from $\vx$ as 
 $\vd_0(\vx)=\phi(\vx,\tau_0(\vx))$ and $\vd_1(\vx)=\phi(\vx,\tau_1(\vx))$. Also, we define the boundary points in terms of the parametrization $s$ for the boundary curve $\vs(s)$: $s_0(\vx), s_1(\vx)$ are defined by
\begin{equation}
\vd(s_0(\vx)) = \vd_0(\vx)  \;\;\;,\;\;\; \vd(s_1(\vx)) = \vd_1(\vx).
\end{equation}
Figure  \ref{fig_2} gives an illustration.
Since $\vv$ is simple, there is a unique $\tau_0,\tau_1,\vd_0,\vd_1,s_0, s_1$ for every $\vx \in D$. Also, (\ref{eq_cond_a_b_4}) gives that it is sufficient to compute $a,b,$ on  the boundary of $D$ only by
\begin{eqnarray}
\label{eq_a_inner}
a(\vx) &=& a(\vd_0(\vx)) = a(\vd_1(\vx)) \\
\label{eq_b_inner}
b(\vx) &=& b(\vd_0(\vx)) - \tau_0(\vx)= b(\vd_1(\vx)) - \tau_1(\vx)
\end{eqnarray}

\subsection{Computing $a$, $b$}

The computation of both $a$ and  $b$ is done in the following steps: 
\begin{enumerate}
\item
Compute $a_s(s) = \frac{d \; a(\vd(s))}{d \; s}$, and  $b_s(s) = \frac{d \; b(\vd(s))}{d \; s}$, i.e., the derivatives of $a,b$ along the boundary curve. 
\item
Compute $a(s) = a(\vd(s))$, $b(s) = b(\vd(s))$ by integrating $a_s(s), b_s(s)$, respectively.
\item
Compute $a(\vx)$, $b(\vx)$ in the inner domain by applying (\ref{eq_a_inner}), (\ref{eq_b_inner}).
\end{enumerate}

\subsection*{Computing $a$}

For $\vx \in D$, we compute the separating function \cite{Friederici:2017:Topo}
along the stream as 
\begin{equation}
\label{eq_comp1}
s_l(\vx,\tau) = \int_{\tau_0(\vx)}^{\tau} \overline{\vw}(\phi)^T  \mJ(\phi) \;  \overline{\vw}(\phi) \; d r
\end{equation}
with $\phi = \phi(\vx,t)$ that describes the behavior of "adjacent" isolines in $a$. We search for the scalar $s_m(\vx)$ minimizing
\begin{equation}
\int_{\tau_0(\vx)}^{\tau_1(\vx)}  \left(e^{-s_l(\vx,\tau) - s_m(\vx)}- \| \vv(\phi(\vx,\tau)) \| \right)^2 d \tau  \to \min.
\end{equation}
Fortunately, this has a closed form solution for $s_m(\vx)$:
\begin{equation}
s_m(\vx) = \ln \left(
\frac{
\int_{\tau_0(\vx)}^{\tau_1(\vx)} (e^{-s_l(\vx,\tau)})^2 d \tau
}{
\int_{\tau_0(\vx)}^{\tau_1(\vx)}    \| \vv(\phi(\vx,\tau)) \|   e^{-s_l(\vx,\tau)} d \tau
}
\right ).
\end{equation}
Finally, we compute 
\begin{equation}
s(\vx) = s_l(\vx,0) + s_m(\vx).
\end{equation}
The field $s(\vx)$ steers the gradient of $a$ by
\begin{equation}
\nabla a =  e^{-s} \overline{\vw}.
\end{equation}

\subsection*{Computing $b$}

If $b$ is set on a stream line, we need to set it on the "adjacent" stream line as well. In other words, we need to study and optimize the behavior of 
$\nabla b$. It turns out that -- similar to $a$ -- we have one degree of freedom to steer $\nabla b$ along the stream line. In the best case we would have 
$(\nabla a)^T \nabla b = 0$. However, if we set this property  in $\vx$, it does not advect along the stream line.

\begin{figure}
  \centering
    \includegraphics[width=0.3 \textwidth]{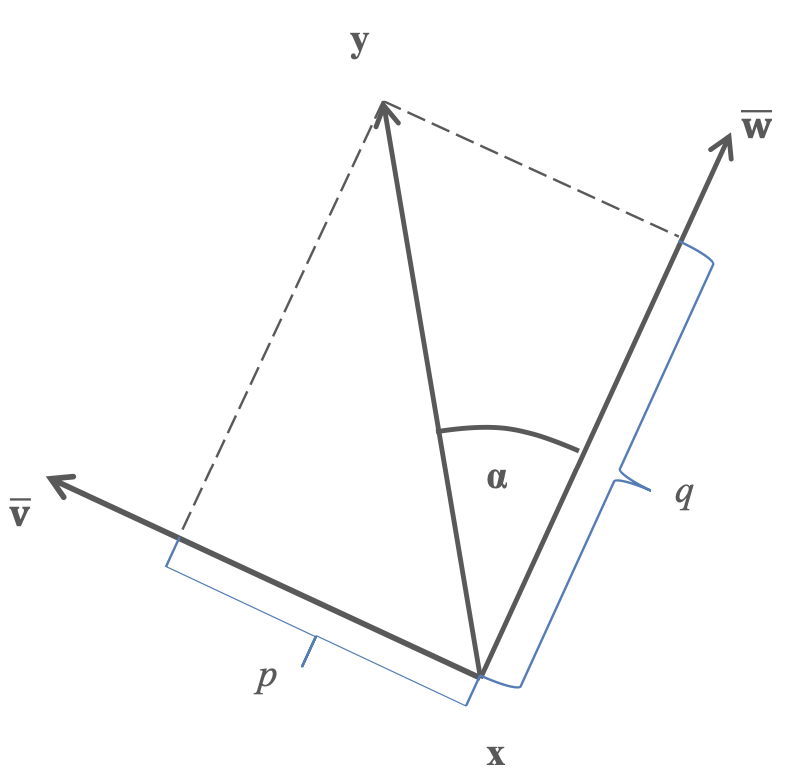}    \hfill
    \caption{setup of local moving coordinate system.}
  \label{fig_3}
 \end{figure}

At a point $\vx$, we consider a point  $\vy$ in its neighborhood and observe $\vy-\vx$ under the local (changing) coordinate system $(\overline{\vv},\overline{\vw})$:
\begin{equation}
p = \overline{\vv}^T (\vy-\vx)
\;,\;
q = \overline{\vw}^T (\vy-\vx)
\;,\;
r = \frac{p}{q} = \tan \alpha.
\end{equation}
Figure   \ref{fig_3} gives an illustration of the setup. 

We assume that  $(\vy-\vx)$ is the direction of the co-gradient of $b$. We observe how, $p,q,r$ behave over stream line integration:
\begin{eqnarray}
\frac{d p}{d \tau} &=&  \overline{\vv}^T \mJ \overline{\vv} \, p + (\overline{\vv}^T \mJ \overline{\vw} + \overline{\vw}^T \mJ \overline{\vv}) \, q \\
\frac{d q}{d \tau} &=& \overline{\vw}^T \mJ \overline{\vw} \, q \\
\label{eq_cond_dr}
\frac{d r}{d \tau} &=&  (\overline{\vv}^T \mJ \overline{\vv} -  \overline{\vw}^T \mJ \overline{\vw})  \, r +  (\overline{\vv}^T \mJ \overline{\vw} + \overline{\vw}^T \mJ \overline{\vv}).
\end{eqnarray}
With (\ref{eq_cond_dr}) we can compute the optimal $r$ along a stream line by 
\begin{equation}
\label{eq_comp2}
\int_{\tau_0(\vx)}^{\tau_1(\vx)} (r(\phi(\vx,\tau)))^2 d \tau \to \min.
\end{equation}
This should be linear in $r$. Note again that $r$ defines the direction of the co-gradient of $b$. Together with (\ref{eq_cond_a_b_3}), $\nabla b$ is uniquely defined.

\section{Gradient-preserving cuts}

Keep in mind that the approach to compute $a$ and $b$ in section
\ref{sec_solution_simple_field} works only on simple vector fields. To make it applicable to every vector field, we introduce gradient-preserving cuts.

\begin{figure}[ht]
\centering
	\includegraphics[width=0.99\columnwidth]{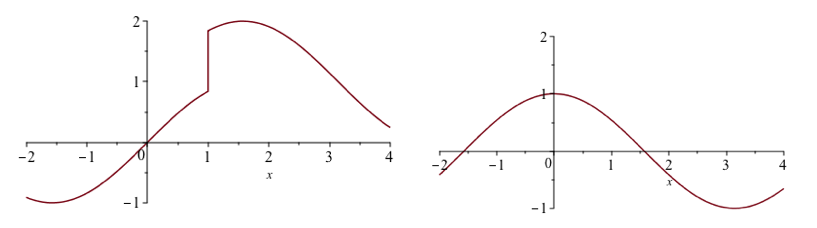}%
	\caption{left: 1D field with gradient-preserving cut; right: its derivative.}
	\label{fig_01}%
\end{figure}

\begin{figure}[ht]
\centering
	\includegraphics[width=0.99\columnwidth]{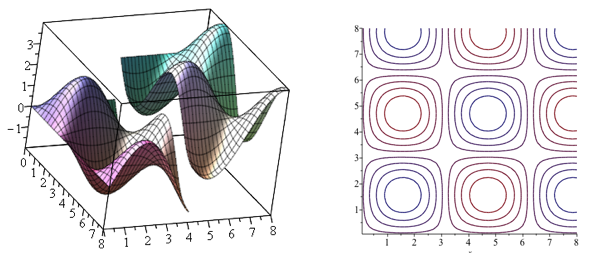}%
	\caption{left: 2D field with gradient-preserving cut as height field; right: isocontours}
	\label{fig_02}%
\end{figure}

In the following we give a definition of gradient-preserving cuts of a 2D scalar field along a cutting curve $\vc(t)$:
 \begin{definition}
 \label{eq_01}
Given is a 2D scalar field $s(x,y)$ and a regularly parametrized curve $\vc(t)$ in the domain of $s$. We assume $s$ to be at least $C^1$ continuous outside $\vc$. The field $s$ has a gradient-preserving cut  along $\vc$ if
\begin{enumerate}
\item
$s$ is undefined on $\vc$.
\item
$\lim_{\vx \to \vc(t),left} s(\vx) - \lim_{\vx \to \vc(t),right} s(\vx) = h$
\item
$\lim_{\vx \to \vc(t),left} \nabla s(\vx) = \lim_{\vx \to \vc(t),right} \nabla s(\vx)$
\end{enumerate}
where $\lim_{\vx \to \vc(t),left}$ considers all points $\vx$ being "left" of the curve, i.e.,  $\det( \dot{\vc}(t) , (\vx - \vc(t) ) > 0$ and $\dot{\vc}$ is the tangent vector of $\vc$. Further, $h$ is a non-zero constant.
 \end{definition}
Definition  \ref{eq_01} states that along $\vc(t)$, $s$ has a well-defined discontinuity but nevertheless a continuous gradient. Figure \ref{fig_01} shows an example of a gradient-preserving cut of a 1D function $s(x)$.
Figure \ref{fig_02} shows a 2D gradient-preserving cut in a scalar field as hight field along with isolines. Note that the isolines appear continuously even though the field has a cut.
Since scalar fields with gradient-preserving cuts are gradient continuous, they have continuous gradient and co-gradient fields. 
Hence, the potential advances still hold under slight modifications of the algorithms: for isocontouring approaches to integrate a co-gradient field, the consideration of an offset $h$ of  isovalues across the cutting curve is necessary. For an integration of a gradient ascent/descent, a special treatment across the cutting curve is necessary.

\section{Treatment of critical points}

\begin{figure}[ht]
\centering
	\includegraphics[width=0.33\columnwidth]{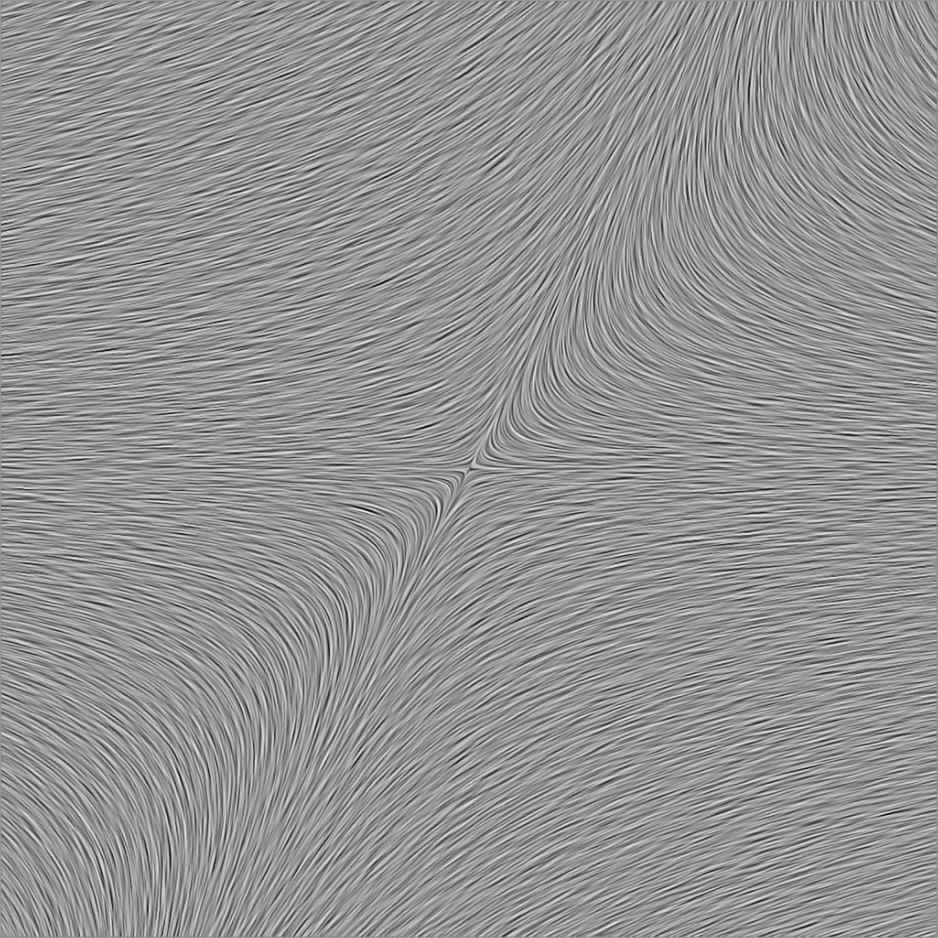}%
	\includegraphics[width=0.33\columnwidth]{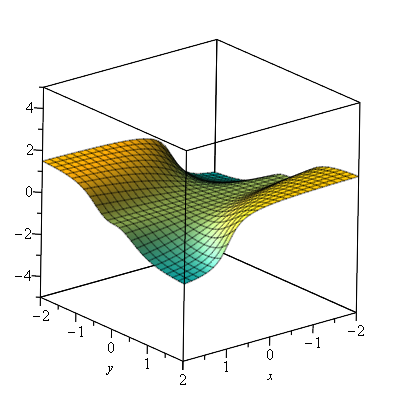}%
	\includegraphics[width=0.33\columnwidth]{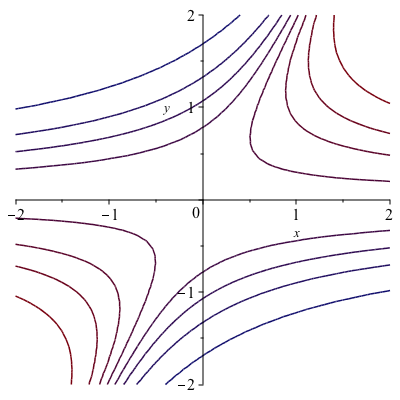}%
	\caption{ 
	Linear vector field $\vv_1$ containing a saddle; LIC image of $\vv_1$ (left), $s$ as height field (middle), isocontours of $s$ (right).
	}
	\label{fig_03}%
\end{figure}

Consider the 2D linear vector field $\vv(\vx) = \mJ \, \vx$ which has a critical point at the origin. Depending on an eigenanalysis of the Jacobian $\mJ$, we discuss the following cases (let $\lambda_1, \lambda_2$ be the eigenvalues of $\mJ$ and $\vr_1, \vr_2$ the corresponding eigenvectors):

\noindent
{\em Case 1:}
$\vv$ has a saddle, i.e., $\lambda_1 < 0 < \lambda_2$. In this case it is a straightforward exercise in algebra to  that the scalar field
\begin{equation}
\label{eq_linear_example_1}
s(\vx) = \det(\vx,\vr_1)^{-\lambda_1}  \det(\vx,\vr_2)^{\lambda_2}
\end{equation}
with $p^q = \mbox{sign}(p) \, \|p\|^q$ fullfills 2., i.e., $\vv^T \nabla s = 0$. Note that (\ref{eq_linear_example_1}) does not even need a gradient-preserving cut to fullfill 2. Figure \ref{fig_03}
illustrates this for the example field 
	$\vv_1 = 
	\left(
	\begin{array}{cc}
	-2 & \frac{3}{2}\\
	0 & 1
	\end{array}
	\right) \vx$ by showing the LIC image of $\vv$ (left), $s$ as height field (middle), and isocontours of $s$ (right).

\begin{figure}[ht]
\centering
	\includegraphics[width=0.33\columnwidth]{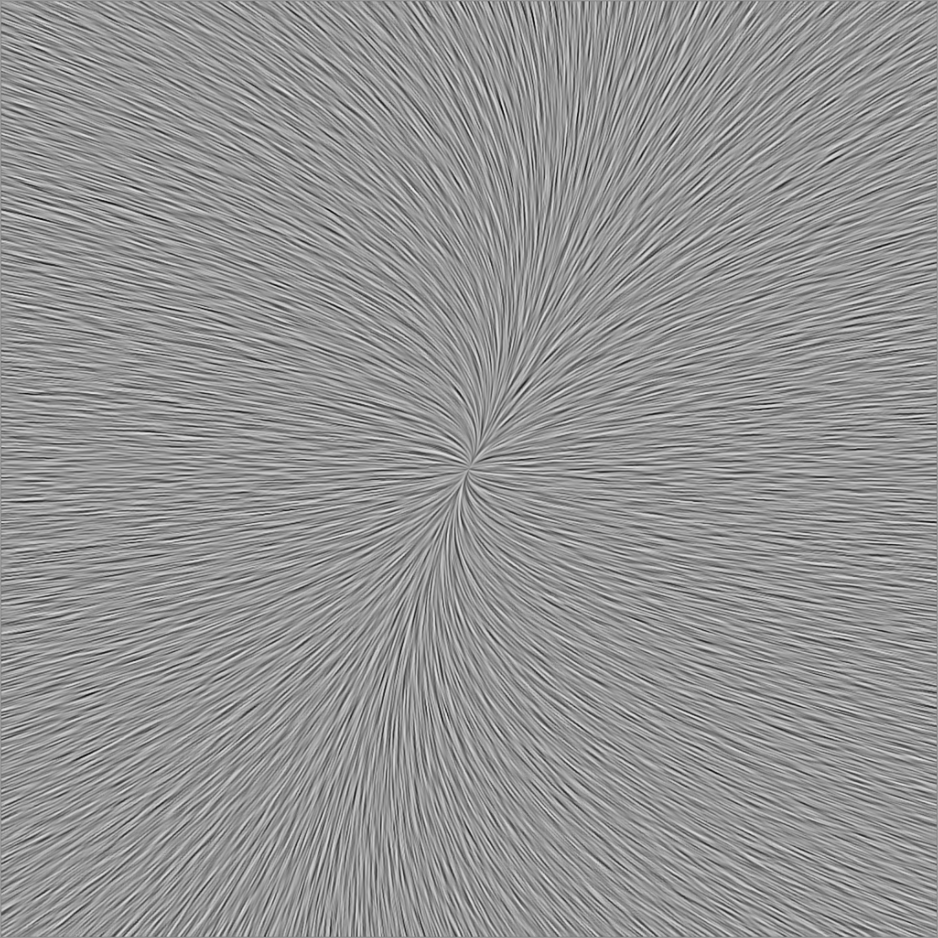}%
	\includegraphics[width=0.33\columnwidth]{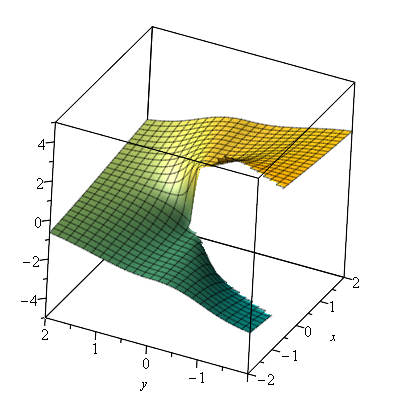}%
	\includegraphics[width=0.33\columnwidth]{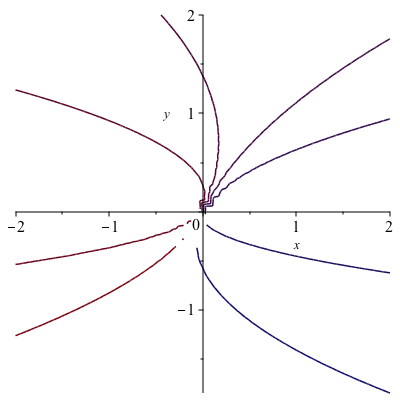}%
	\caption{Linear vector field $\vv_2$ with two positive real eigenvalues of $\mJ$;  LIC image of $\vv_2$ (left), $s$ as height field (middle), isocontours of $s$ (right).
	}
	\label{fig_04}%
\end{figure}

\noindent
{\em Case 2:}
$\vv$ has a source or a sink with real eigenvalues of $\vv$, i.e., $\lambda_1  \leq \lambda_2 < 0$ or $0 < \lambda_1  \leq \lambda_2$. In this case the scalar field
 \begin{equation}
\label{eq_linear_example_2}
s(\vx) =   \arctan \left(  \det(\vx,\vr_1)^{-\lambda_1}  \det(\vx,\vr_2)^{\lambda_2} \right)
\end{equation}
fulfills 2. Moreover, here we have a gradient-preserving cut starting from the origin that is due to the arctan function. Figure \ref{fig_04} illustrates this for the vector field
	$\vv_2 = 
	\left(
	\begin{array}{cc}
	2 & -\frac{1}{2}\\
	0 & 1
	\end{array}
	\right) \vx$.

\noindent
{\em Case 3:}
$\vv$ is a source or a sink with imaginary eigenvalues of $\mJ$, i.e., a swirling behavior around the critical point. In this case, the scalar field 
\begin{equation}
\label{eq_linear_example_3}
s(\vx) =   \alpha \arctan \left( \frac{\va^T \vx }{\beta \, \vb^T  \vx} \right) + \beta \, \ln(\vx^T \,\mC\, \vx)
\end{equation}
with
\begin{eqnarray}
\nonumber
\mR = 
	\left(
	\begin{array}{cc}
	0 & -1\\
	1 & 0
	\end{array}
	\right)  
\;\;\;,\;\;\;
\widehat{\mJ} = \frac{1}{2}  \left(  \mR \, \mJ + (\mR \, \mJ)^T ) \right)
\;\;\;,\;\;\;
\alpha = \mbox{Trace}(\mJ)
\\
\nonumber
\beta = \sqrt{  \|   \det \widehat{\mJ} \|  }
\;\;\;,\;\;\;
\va^T = (1,0) \, \widehat{\mJ}
\;\;\;,\;\;\;
\vb^T = (1,0) \, \mR
\;\;\;,\;\;\;
\mC = \widehat{\mJ}.
\end{eqnarray}

\begin{figure}[ht]
\centering
	\includegraphics[width=0.33\columnwidth]{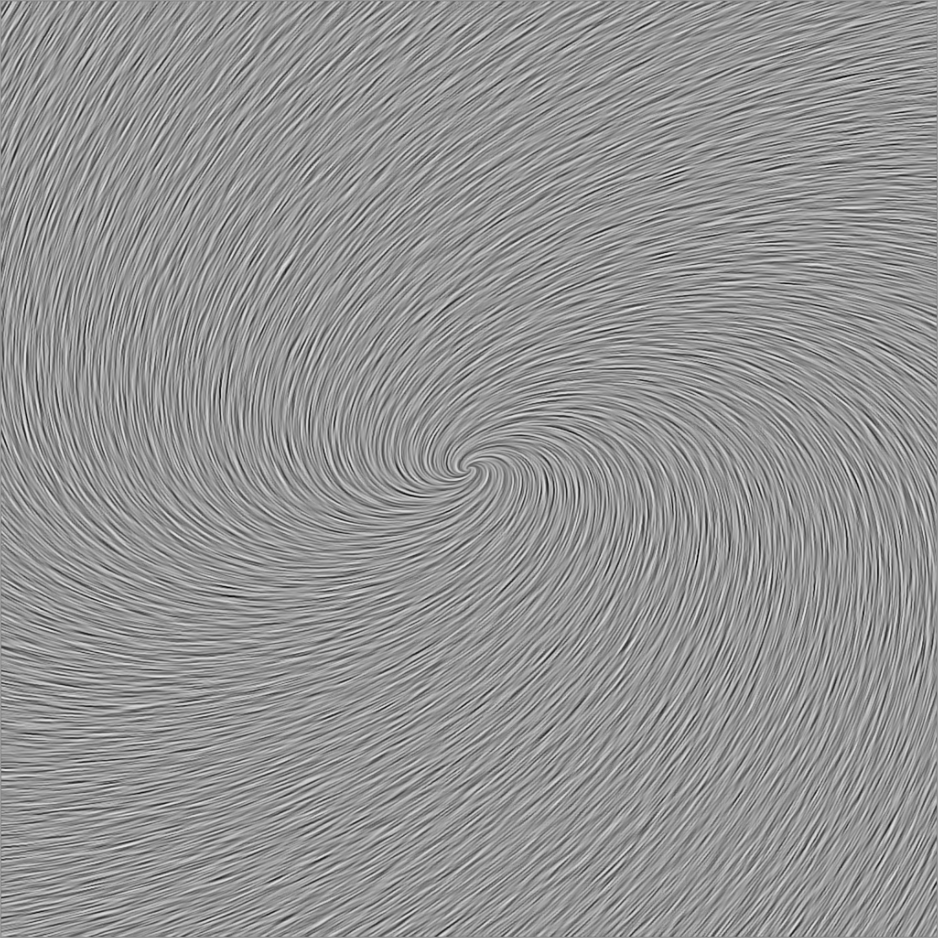}%
	\includegraphics[width=0.33\columnwidth]{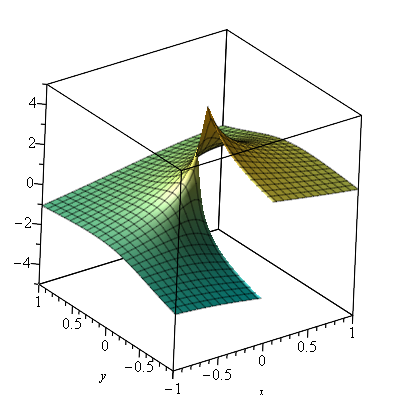}%
	\includegraphics[width=0.33\columnwidth]{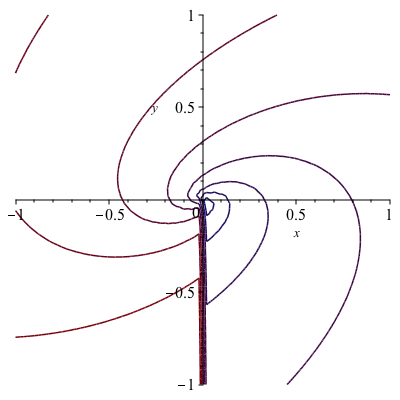}%
	\caption{Vector field $\vv_3$ with imaginary eigenvalues of the Jacobian; LIC image of (left), $s$ as height field (middle), isocontours of $s$ (right).
	}
	\label{fig_05}%
\end{figure}

Figure \ref{fig_05} illustrates this for the field 
	$\vv_3 = 
	\left(
	\begin{array}{cc}
	1 & 4\\
	-2 & 3
	\end{array}
	\right) \vx$.

\noindent	
Based on these cases, we formulate the following: 
\begin{corollary}
Every 2D linear vector field can be described as co-gradient field of a scalar field $a$ with a gradient-preserving cut.
\end{corollary}

\section{Placement of the cuts}

Cuts must be placed such that all critical points are covered and the vector field $\vv$ is divided into a set of simple vector fields. To this end, we set the cuts manually by defining a polygon connecting all critical points and to connect to one additional point at the boundary of the domain. A similar approach for defining cuts has been applied in \cite{Theisel:2004:VMV} to compute isolated closed stream lines. 

\section{Discretization and solution}

We assume a piecewise linear vector field over a triangulation of the domain $D$. We search for piecewise scalar fields $a$, $b$ over the same triangulation, i.e., the unknown scalar values $a$, $b$ at the vertices of the triangulation. The general algorithm is as follows:
\begin{enumerate}
\item
Set the cuts in the domain $D$.
\item
Compute $a$, $b$ along the boundary of $D$, i.e., 
\begin{equation}
a(s) = a(\vd(s))  \;\;\;,\;\;\; b(s) = b(\vd(b)). 
\end{equation}
\item
For every inner point $\vx$ on the triangulation, compute $a(\vx)$ and  $b(\vx)$ by applying (\ref{eq_a_inner}), (\ref{eq_b_inner}).
\end{enumerate}
More information is necessary for step 2. We compute all boundary switch points \cite{Weinkauf:2004:VisSym} on the boundary of $D$ that divide the boundary in alternating areas of inflow and outflow, respectively. We consider outflow areas only and compute $a$, $b$ by applying (\ref{eq_comp1})--(\ref{eq_comp2}). For inflow areas, the corresponding values  $a$, $b$ are computed by applying (\ref{eq_a_inner}), (\ref{eq_b_inner}), respectively.

\section{Results}

\begin{figure}[ht]
\centering
	\includegraphics[width=0.5\columnwidth]{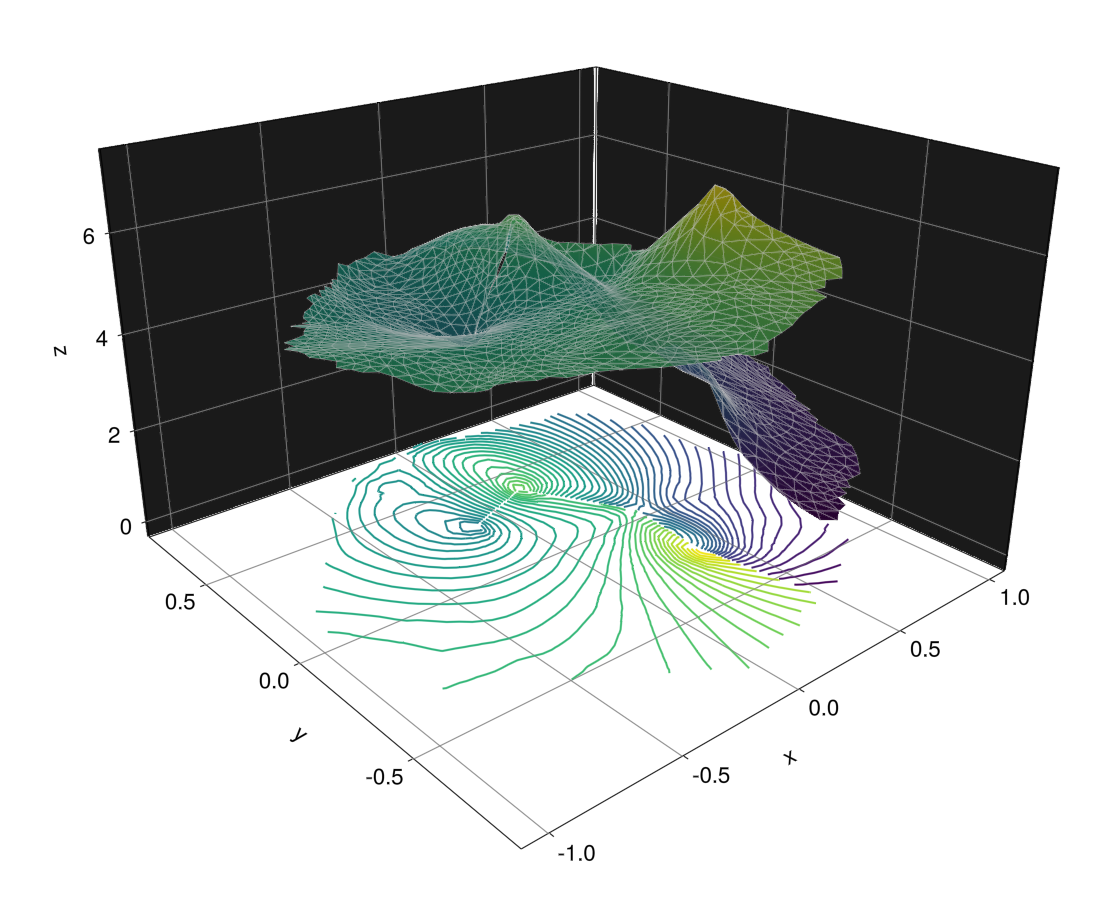}%
	\includegraphics[width=0.5\columnwidth]{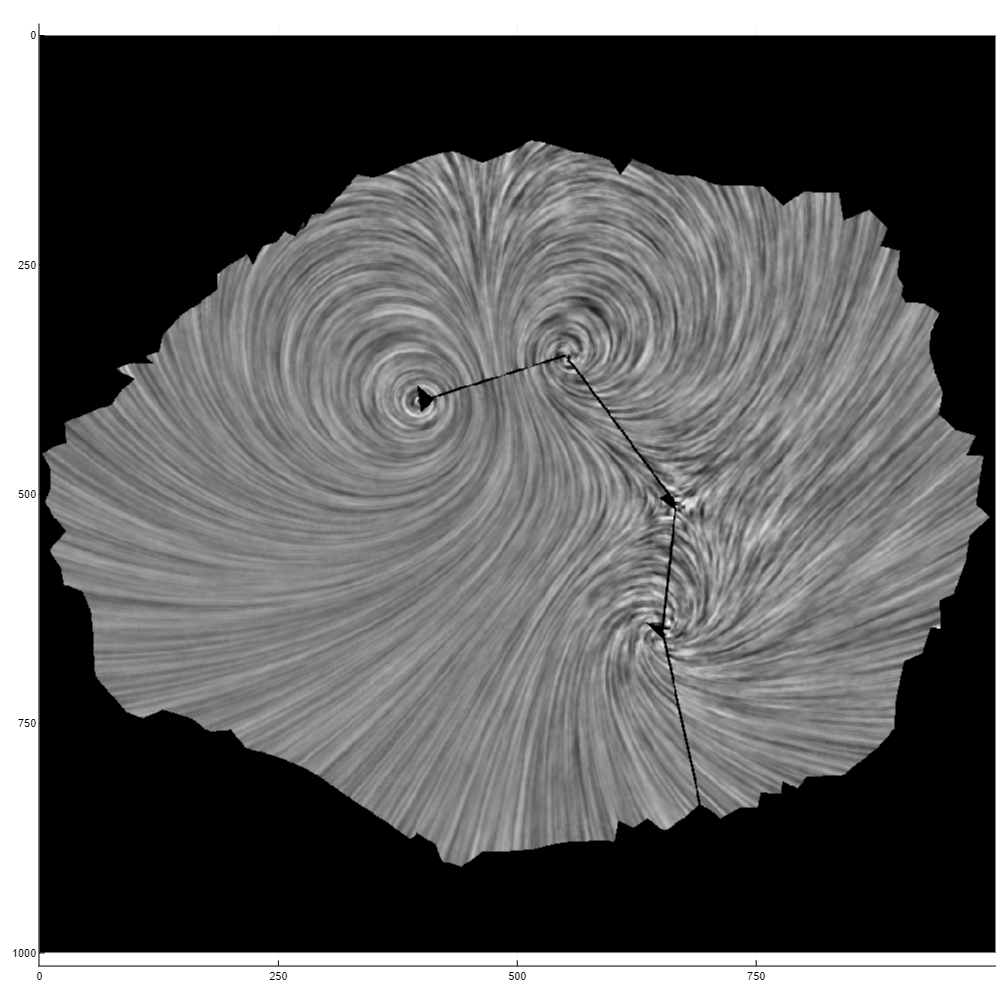}%
	\caption{Vector field with 3 swirling sources and a saddle; 3d image of the constructed scalar field (left), LIC of the vector field (right).
	}
	\label{fig_10}%
\end{figure}

We apply our approach to a number of test fields. Figure \ref{fig_10} shows a piecewise linear vector field with 3 swirling sources and a saddle. 
Figure \ref{fig_10}(right) shows a LIC image of the field as well as the cuts. Figure \ref{fig_10}(left) shows the computed piecewise linear field $a$ as height field. Also the underlying triangulation for the piecewise linear fields are shown. The projected lines in figure  \ref{fig_10}(left) show isolines of $a$ that correspond to stream lines of $\vv$. Note that we can see a clear discontinuity of $a$ at the cuts, but no discontinuities are visible in the isolines, which is due to the gradient-preservation property.   

\begin{figure}[ht]
\centering
	\includegraphics[width=0.5\columnwidth]{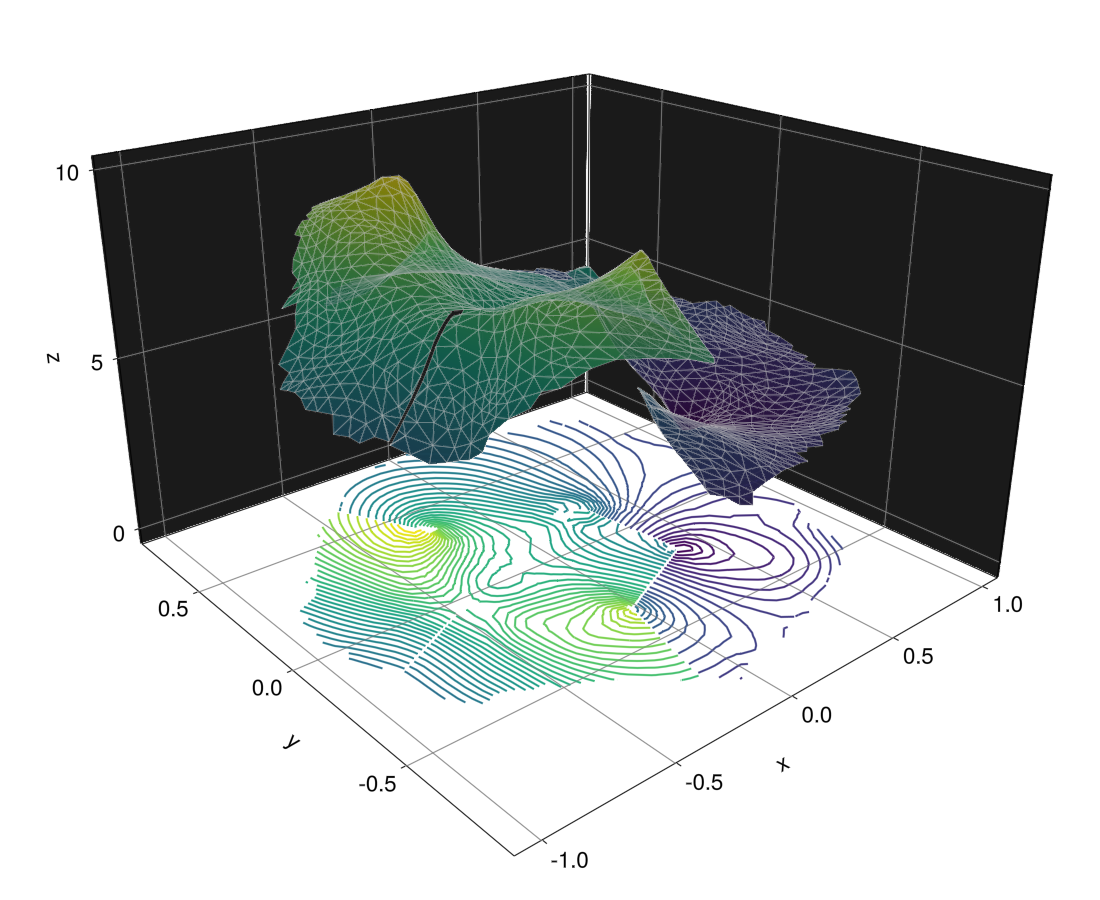}%
	\includegraphics[width=0.5\columnwidth]{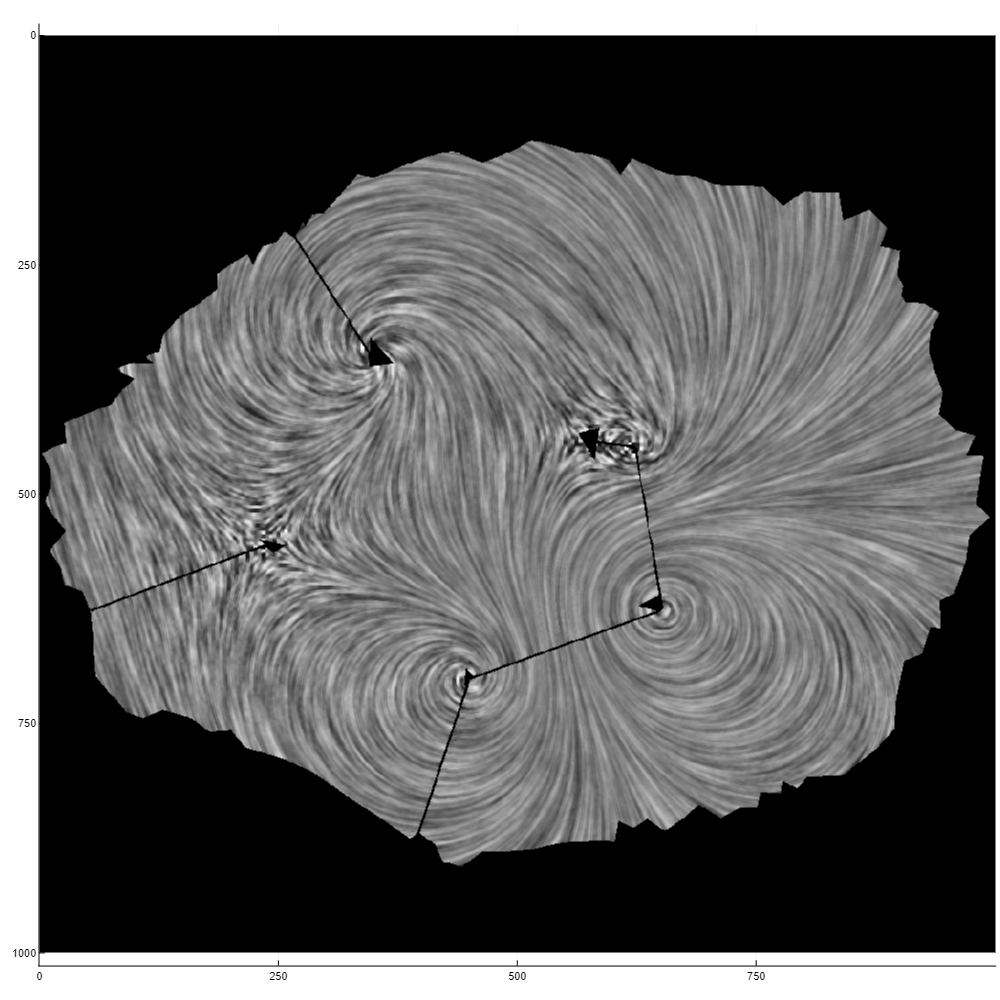}%
	\caption{Vector field with 2 sources, 2 sinks and 2 saddles; 3d image of the constructed scalar field (left), LIC of the vector field (right).
	}
	\label{fig_11}%
\end{figure}

Figure \ref{fig_11} shows the approach for a vector field with 2 sources, 2 sinks and 2 saddles.

\section{Conclusion}
 
We presented a principal solution to represent 2D vector fields as scalar fields. Further research is necessary concerning the comparison of accuracy of isoline extraction of $a$ vs. a standard numerical stream line integration of $\vv$.  Also, further and more complex datas sets need to be evaluated.
 
An extension to 2D time-dependent as well as to 3D vector fields is non-trivial, but there is no fundamental reason that prevents the extension. We leave this extension to future research.

\section*{Acknowledgements}

This work was partially supported by DFG grant TH 692/17-1.

\bibliographystyle{unsrtnat}
\bibliography{vectorfieldasscalarfield}  






\end{document}